# Self-Censorship Under Law:
# A Case Study of the Hong Kong National Security Law

Mona Wang
*Princeton University*

Jonathan Mayer
*Princeton University*

## Abstract

We study how legislation that restricts speech can induce online self-censorship and alter online discourse, using the recent Hong Kong national security law as a case study. We collect a dataset of 7 million historical Tweets from Hong Kong users, supplemented with historical snapshots of Tweet streams collected by other researchers. We compare online activity before and after enactment of the national security law, and we find that Hong Kong users demonstrate two types of self-censorship. First, Hong Kong users are more likely than a control group, sampled randomly from historical snapshots of Tweet streams, to remove past online activity. Specifically, Hong Kong users are over a third more likely than the control group to delete or restrict their account and over twice as likely to delete past posts. Second, we find that Hong Kong users post less often about politically sensitive topics that have been censored on social media in mainland China. This trend continues to increase.

## 1  Introduction

Empirical research about censorship of online speech focuses on measuring network- and platform-level content restrictions. Governments do often censor online speech by outright preventing access to content, but there is another important form of censorship that has received far less quantitative study: self-censorship, when a government chills online speech by imposing legal (and sometimes extralegal) risks. Despite a wealth of qualitative research, media reporting, and public writing about self-censorship, there have been relatively fewer large-scale quantitative case studies that demonstrate the effect of speech law on online self-censorship at scale.

On June 30, 2020, the Chinese National People's Congress enacted a "national security law" (NSL) with direct application to Hong Kong [27]. The NSL entered into force the same day. The new law criminalizes speech that the government deems "seditious" or "secessionist" in nature, terms which the law implements with broad and ambiguous prohibitions.

Anecdotally, individuals and organizations in Hong Kong have recently curtailed their own activities and censored their own speech both offline and online, especially after the national security law (NSL) entered into force. Users on LIHKG, a social media site popular in Hong Kong, suggested that Hong Kongers delete social media accounts or past social media activity that could incriminate them under the NSL [7]. Since the NSL's enactment, Hong Kong authorities have arrested people on the basis of their online activity, and local lawmakers have proposed additional bills to outlaw specific forms of online speech (e.g., "Publishing an image of a defiled national flag on Facebook") [6, 9].

Our work aims to demonstrate the impact that newly enacted liability can have on online speech, using Hong Kong's national security law as a case study. We quantify at scale the self-censorship exhibited by social media users after the national security law passed. We use Twitter datasets because, according to social media market research, about 29% of Hong Kongers aged 19-64 used Twitter in 2021 [21]. Our work seeks to answer two research questions.

**RQ1**: Comparing social media activity by Hong Kong users before and after enactment of the national security law, how often do Hong Kong users delete posts or accounts, and how does the frequency compare to other groups of users?

**RQ2**: Comparing social media activity by Hong Kong users before and after enactment of the national security law, how has the amount of discussion of sensitive political topics possibly covered by the NSL changed relative to other topics, and how does this discussion compare to a control group?

To answer these research questions, we curate several datasets of Tweets and Twitter users. First, we compile a dataset of archived social media data from before the NSL entered into force. This historical dataset includes posts and accounts that may have subsequently been deleted or made private. The archived data consists of about 2 million Tweets from various user populations during 2019. Second, we compile datasets of currently available Tweets and Twitter users from before and after enactment of the NSL. These datasets contain over 7 million Tweets from Hong Kong users and 8 million Tweets from a set of control users.

In our analysis to answer **RQ1**, we find that Hong Kong users are over a third more likely than a control sample to protect their accounts and over twice as likely to delete past Tweets than control Twitter users.

To address **RQ2**, we additionally curate a dataset of Tweet keywords that were common among Hong Kong users before the NSL and that are associated with politically sensitive topics that are censored on social media platforms in mainland China. We analyze the relative frequency of Tweets containing politically sensitive keywords over time for Hong Kong users and for a control group. We find that Hong Kong users continue to speak less online about politically sensitive topics.

Our case study presents large-scale quantitative evidence that aggressive legislation and policy can quickly and starkly alter the nature of online political discourse.

## 2 Background

In this section we present prior research measuring self-censorship in online discourse, and we offer background for our Hong Kong case study.

### 2.1 Measuring self-censorship in online discourse

We define self-censorship, or the chilling effect, consistent with prior scholarship: when an individual withholds or falsifies discourse for fear of repercussion [20]. There is a vast literature on measuring online political discourse [11, 13, 25]. There is also a large body of qualitative research, especially in law, public policy, and politics, about self-censorship and chilling effects [14, 19, 20]. The media also often reports on this phenomenon, often from anecdotal evidence or hypotheses by policymakers [16–18]. There is, however, very little large-scale empirical research on changes in online political discourse that are attributable to self-censorship.

Past research has shown that measurable differences can surface around discrete events that increase the perception of online surveillance. In the most similar prior work, Tanash et al. quantified the change in Tweeting behavior by Turkish users specifically after the 2016 attempted coup in Turkey [23]. The Turkish government subsequently arrested thousands of people that it blamed for plotting the coup, with little due process. Many of these arrests resulted from investigations into social media activity, solely on the basis of individuals' online speech and actions. Notably, Tanash et al. measured both a surge in retroactively deleted tweets by Turkish users and a significant decrease in certain politically sensitive tweets from Turkish accounts.

### 2.2 The Hong Kong national security law

The new national security law for Hong Kong creates penalties for people who participate in secession, subversion of the governments of mainland China or Hong Kong, terrorist activities, or collusion with a foreign country to endanger national security. In addition to having a vague and sweeping scope, the law extends beyond Hong Kong: Article 38 establishes liability for offenses that occur "outside the region by a person who is not a permanent resident of the region" [27].

In the six months after the NSL entered into force, Hong Kong law enforcement arrested at least 100 individuals on the basis of the new law [26]. At least 24 of the arrests involved charges related to "seditious" or "secessionist" speech. The arrestees included legislators, protestors, student activists, journalists, and an American human rights lawyer. Journalists in Hong Kong have described their fear of declining press freedoms and increased self-censorship in the media [3]. Because enforcement of the NSL has already targeted online political speech, Hong Kongers may have a strong incentive to self-censor their online social media activity.

## 3 Methodology

In this section, we describe how we collect data to answer our two research questions. For each, we curate several large datasets, and we perform various analyses on the data.

To curate these datasets, we combine various sources of Twitter data both from archives and from Twitter's Full-Archive Search API. We then augment the data with additional live data from the Twitter API. Next, we filter and curate these large data sources into smaller datasets, which we use for analysis. We enumerate our datasets in Table 1.

Figure 1 illustrates the data collection process for our study. The code for the data collection and analysis can be found at https://github.com/citp/hk-twitter.

### 3.1 Post and account deletion

**RQ1**: Comparing social media activity by Hong Kong users before and after enactment of the national secu-rity law, how often do Hong Kong users delete posts or accounts, and how does the frequency compare to other groups of users?

|   | Size | Date range | Snapshot date | Source |
|---|---|---|---|---|
| A. Tweets geotagged in HK | 4M | [2018/1, 2020/12] | 2021/4 | Twitter API |
| B. Tweets by HK users in Tweet stream archives | 260K | [2019/1, 2019/12] | [2019/1, 2019/12] | Archived data |
| C. HK user dataset | 36K | N/A | 2021/4, 2021/12 | A and B |
| D. Tweets by 1.5k HK users | 7M | [2019/1, 2022/05] | 2022/08 | C and API |

Table 1: A summary of the Tweet and Twitter user datasets that we used for analysis. *Date range* indicates the dates that the Tweets were published, and *snapshot date* is when the Tweets were retrieved. We note that for datasets A, B, and C, we also collected equivalent datasets for New York, Taipei, and Tokyo for comparison. We also collected a randomly sampled control dataset from the Tweet stream archives used in B, and extracted equivalent user and Tweet datasets for C and D using the control sample.

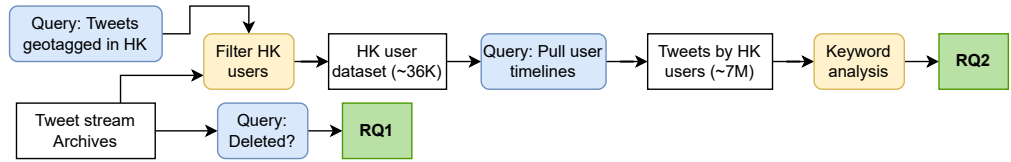

Figure 1: Main experiment data pipelines. The blue "Query" boxes access the Twitter v2 API. To generate the control and datasets from other locations, we change the "Filter HK users" step to either sample from users uniformly at random (for the control set) or to filter users with profiles in different cities or geographical areas.

To answer this question, we must rely primarily on historical data sources. In particular, we wish to identify Tweets made by Hong Kong users before the national security law passed and then query whether those Tweets and the associated users are still available on Twitter. We compare deletion rates across Hong Kong, Tokyo, Taipei, and New York City, and we also compare removal rates to those in a randomly sampled control dataset.

We use the Twitter v2 API to query whether a user account has been protected (i.e., is private and only visible to followers) or has been deleted. For accounts that are still public, we query whether their Tweets in the archived data have been deleted since they were originally posted.

### 3.1.1 Dataset sources

We use several publicly available Twitter archives to curate a historical Tweet dataset.

**Pushshift.** Pushshift is a social media data collection, analysis, and archiving platform that archives social media posts from a variety of social media sites, including Reddit and Twitter [4,5]. Their Twitter archives contain two snapshots of Hong Kong-related queries to Twitter's Recent Search, collected in August of 2019. This collection contains a total of 380k Tweets containing the Anti-Extradition hashtag, from August 2019 during the peak of protests against the NSL. Pushshift also collected all Tweets by verified users between May 2019 to November 2019.

**Internet Archive.** Archive Team, a group of social media researchers, has collected data from Twitter's Spritzer Stream (which live samples 1% of all Tweets) from 2012 through part of 2021 and published this dataset on the Internet Archive [2]. While some months are missing from the archive, on the whole this Twitter Stream snapshot provides a consistent large-scale source of historical Tweet data.

### 3.1.2 Dataset curation

From these Twitter stream snapshots in 2019, we extract datasets of Tweets and Twitter users from various locations.

To compare user activity across locales, we must associate the users in our dataset with the areas in our study. We accomplish this by relying on the user's account location, which is self-reported and optional text that appears on a user's account. For instance, we check whether the account location includes *hong kong* or a cognate (e.g., 香港, 🇭🇰, or HK), as determined by the GeoNames database. We use this database since it is also used by the Twitter API to determine profile location [10]. We forgo other signals of user location due to methodology limitations.

**Other potential signals of user location.** Van der Veen et al. found that the two strongest signals of user location were the user's self-reported location and time zone data [24]. This information has previously been used to infer location in other studies of Hong Kong Twitter populations [22].

Time zone data is no longer available from the Twitter API as of 2017 [15]. Twitter now considers a time zone to be private since it is possible to learn information about a user's location from their time zone settings. It may be possible to estimate a user's time zone from the times that they Tweet. However, drawing inferences from Tweet timing may not be feasible for the components of this project that involve deleted or restricted accounts. The method could require more information about a Twitter user's past Tweets than is available in archived data.

We also considered using more fine-grained language identification as a signal for identifying user location. Twitter's automated language identification does not distinguish, for instance, between simplified Chinese, traditional Chinese, and written Cantonese, and often does not perform well distinguishing Japanese from these. Social media users in East Asia often use several different languages, which further confounds this approach to associating users with locations. In our datasets of Tweets by users in Hong Kong, more Tweets contained Japanese characters than traditional Chinese characters.

**Selection of cities for additional dataset curation.** We curate additional datasets for similarly sized metropolitan areas by using an analogous methodology. We take this step because a randomly-sampled control group of users may not account for correlates of online activity that are independent of the NSL. In particular, the cultural and linguistic context of Twitter users may cause Twitter to vary in topics of discussion between different regions. The differences could be especially large between Twitter users generally inside or generally outside East Asia. We chose Taipei and Tokyo due to the popularity of Twitter in those metropolitan areas, their physical proximity, their composition of Tweet languages, and their presence in East Asia. We also choose New York City as a similarly-sized metropolitan area with a markedly different cultural and linguistic context for comparison.

**Control dataset.** We also curate a control dataset. To control for other unknown variables by selecting only accounts with a clearly defined self-reported location (for instance, selecting fewer bot accounts), we first filter our data sources for Tweets that belong to users with a non-empty self-reported location. Then, we sample Tweets uniformly at random from the Twitter stream snapshots to create our control dataset. These Tweets are sampled from the same period of time as the other datasets (i.e., the year 2019).

We note that for the rest of the paper, any reference to the "control" dataset refers to this one. We explicitly name the additional datasets from other metropolitan areas when appropriate.

## 3.2 Changes in online discussion trends

**RQ2**: Comparing social media activity by Hong Kong users before and after enactment of the national security law, how has the amount of discussion of sensitive political topics possibly covered by the NSL changed relative to other topics, and how does this discussion compare to a control group?

To answer this question, we first curate a collection of Tweets by Hong Kong users, then perform a time series bag-of-words analysis on the data, using collections of politically sensitive keywords that may be considered "seditious" or "secessionist" under the NSL.

We curate our dataset of Tweets by fetching the timelines of 1.5K Hong Kong users sampled from our larger Hong Kong user dataset from Section 3.1.2. We identify particular popular keywords that correspond with potentially politically sensitive topics that may be considered criminal under the national security law. We perform the same analysis on a set of 1.5K randomly sampled control users.

### 3.2.1 Dataset curation

To answer the second research question, we require a larger sample of Tweets from users in Hong Kong from a period of time before and after the NSL was passed.

To supplement our datasets from Section 3.1.2, we applied for Twitter's Academic Research API. This API allows us to use Twitter's Full-Archive Search API to query historical Tweets. It also gives us a much larger Tweet quota of 10 million Tweets per month. We used Full-Archive Search to query all Tweets geotagged in Hong Kong over the past several years, as well as to pull all Tweets of users whose profiles were listed in Hong Kong.

**Tweets geotagged in Hong Kong 2018-2020.** Via the Twitter API, we obtained all Tweets (excluding Retweets) that were geotagged in Hong Kong in 2018, 2019, and 2020. We performed this query in April 2021, and the dataset includes approximately 4 million Tweets.

We filter the dataset of Tweets geotagged in Hong Kong for users that are self-reported to be in Hong Kong. Through manual evaluation of 100 random Tweets before and after this filtering process, we find that this process removes many of the most prominent promotion and marketing accounts.

**Tweets by 1,500 Hong Kong users in 2019-2020.** From the Hong Kong user dataset described in Section 3.1.2, we first filter for accounts that Tweet under 50 times per day on average, using the Tweet counts API. We perform this filtering for two reasons. First, we are primarily interested in Tweets by individuals rather than organizations. From a manual review of 30 organizational, promotional, and marketing accounts in our dataset, they Tweeted significantly more frequently. Second, we have a limited Tweet quota, and pulling the full historical archives for high-frequency Tweeters would both consume our quota quickly and skew our dataset towards high-frequency Tweeters. After filtering by Tweet frequency, we sample 1,500 users uniformly at random who currently have self-reported locations in Hong Kong and we fetch all their Tweets in 2019, 2020, 2021, and the first half of 2022. This dataset contains approximately 7 million Tweets. We queried this data in August of 2022, so any accounts or Tweets that were deleted or protected before then will not appear in this dataset.

**Tweets by 1,500 control users in 2019-2020.** We perform the exact same filtering and data collection as above for a sample of our randomly sampled control users from Section 3.1.2. This dataset contains 8 million Tweets.

### 3.2.2 Dataset analysis

We perform a time series keyword matching analysis on Tweets by Hong Kong users, using collections of politically sensitive keywords that may be considered "seditious" or "secessionist" under the NSL. A key component of this study is to curate keyword lists that identify sensitive political speech as opposed to other speech. We expect the former to decrease over time around when the NSL entered into force and the latter to stay constant.

**Curating keyword lists.** We curate three keyword lists: the first contains "sensitive" political keywords as determined by the NSL, "non-sensitive" keywords, and COVID-19 keywords.

Our keywords of "sensitive" political topics consists of terms that researchers at Citizen Lab have previously identified as commonly censored for political reasons by chat applications in mainland China [12]. We include translations of each term in the most popular written languages in our Tweet dataset: English, simplified Chinese, traditional Chinese, and Japanese using machine-assisted translation with verification from native speakers.

We augment this list by manually selecting semantically similar keywords from the 1,000 most popular hashtags in dataset of Tweets by Hong Kong users. For instance, from the top 10 hashtags, we considered keywords "HongKong", "CCP", and "China", as potentially sensitive. The remaining keywords from the top 10 hashtags were about popular multiplayer games, music groups, or other media, which we classified as non-sensitive. We note that we cannot calibrate this list well to our control dataset, since these topics are discussed less often outside Hong Kong (and thus do not appear in popular Twitter trends), and may not be considered "sensitive" outside the political context of the NSL.

We then curate our list of "non-sensitive" keywords. We extract the 1,000 popular trends, this time from both our Tweets by Hong Kong users and Tweets by control users datasets. We again semantically compare the topics to Citizen Lab's keyword blocklist to manually curate collections of keywords that are likely not sensitive, and do not fall under the umbrella of the NSL. The vast majority of these topics concern video games, pop groups or stars, shows, and other popular media.

We also curate a collection of terms concerning COVID-19, using COVID-related keyword blocklists collected by Citizen Lab from Chinese chat applications [12]. Though certain COVID-19 discussion has been censored in mainland China, we were uncertain whether COVID-19 discussion would be considered "sensitive" even under the NSL. Nonetheless, it is a political topic with a huge amount of Twitter discussion both inside and outside Hong Kong and worthy of analysis. We might expect online discussion to peak during earlier waves of COVID-19 transmission both globally and within Hong Kong.

To validate that our keyword lists capture the types of Tweets we are concerned about, we use a simple manual validation method. We randomly sample 100 Tweets in for each keyword group from our Hong Kong dataset as identified by keyword matching (not politically sensitive, politically sensitive, and COVID-related), and we manually review whether the Tweet constitutes political or COVID discussion. We recruited manual reviewers fluent in the languages that were present in these samples, including English, Mandarin Chinese, Cantonese, Japanese, Korean, Thai, and Tagalog. In the non-sensitive sample, reviewers classified 2 of 100 Tweets as discussing geopolitics relating to Hong Kong. While this is a low false negative rate for "sensitive" political Tweets, there are many more Tweets in this category, so we may miss a large portion of "sensitive" political Tweets. In the COVID sample, reviewers classified 2 of 100 Tweets as not discussing COVID-19. In the political sample, reviewers classified 7 of 100 Tweets as not discussing geopolitics relating to Hong Kong. While this keyword-based classification is simple and may have low recall, this classification performance should be stable for an aggregate analysis of trends over a period of time.

**Limitations.** We note biases in the data due to our sampling methods. Not every Tweet is necessarily authored by an individual. Businesses often use geo-tagged Tweets as a marketing tactic to promote various hashtags and topics as local *Trending* topics. We note that the largest Tweeters in the geotagged dataset are lightly automated marketing or promotion accounts. Geotagged Tweets by individuals are also much more likely to come from users on mobile devices.

Excluding high-frequency Tweeters from the Tweets datasets described in Section 3.2.1 due to API limitations introduces further bias to our data. Though we sample our control dataset in the same fashion, this means we cannot draw conclusions about accounts that Tweet at a consistently high frequency.

In addition, while we characterize trends that occur surrounding enactment of the NSL, and these trends appear attributable to the NSL, we do not isolate other possible causal factors in this study. For instance, the Hong Kong anti-extradition protests in 2019 may have prompted greater political discussion by Hong Kong users on Twitter, which our methods would capture as heightened discourse about sensitive political topics in the period before the NSL.

We also note that the "sensitive" keyword list is calibrated specifically to the context of Hong Kong's national security law. These topics did not occur in the top 1,000 hashtags for the control dataset, and in fact are discussed an order of magnitude less often globally than in Hong Kong. This difference in calibrations may also introduce bias. At the very least, the relative frequency of "sensitive" keywords in the control data will be noisier than in the Hong Kong data since these topics are discussed less frequently.

Finally, we do not conduct sentiment analysis on these posts or other finer-grained analysis. This work observes overall discussion about "sensitive" political topics in aggregate and does not differentiate between self-censorship of particular viewpoints on these topics. We still expect overall discussion to steadily decrease after the national security law passed; as observed in prior work, self-censorship induced by local policy can stifle overall speech [23].

## 4 Results

In this section, we discuss the results from our experiments. First, we discuss RQ1, specifically post and account deletion rates of Hong Kong users compared to other groups of users. Then, we discuss RQ2, and identify changes in online discussion trends of Hong Kong users around the time the NSL was passed.

|  | HK | NYC | Tokyo | Taipei | Control |
|---|---|---|---|---|---|
| # accounts | 35.9K | 18.0K | 215K | 39.8K | 128K |
| ...% protected | 8.86% | 5.45% | 7.29% | 11.1% | 6.62% |
| ...% deleted | 14.9% | 10.4% | 14.3% | 11.9% | 14.3% |
| # Tweets | 264K | 133K | 676K | 243K | 130K |
| ...% deleted | 34.1% | 28.6% | 27.9% | 33.9% | 15.0% |
| ...% inaccess. | 55.76% | 45.13% | 45.93% | 47.54% | 45.67% |

Table 2: Deletion rates of user accounts and Tweets across different self-reported locations. The Tweets occur before the NSL, in the year 2019. We then query the live Twitter API in 2022 to identify whether these Tweets or users are still accessible. Control takes a random sample of users. % Tweets deleted is the percentage of Tweets belonging to public accounts that have been deleted. % inaccessible also includes Tweets belonging to users that were protected or deleted.

|  | HK | NYC | Tokyo | Taipei |
|---|---|---|---|---|
| # sensitive Tweets | 12.2K | 0.7K | 5.1K | 4.6K |
| ...% inaccess. | 36.38% | 29.45% | 30.80% | 34.89% |

Table 3: Rate of Tweet inaccessibility for Tweets containing keywords that are related to those commonly censored on Chinese chat apps.

### 4.1 RQ1: Post and account deletion

Our results for analyzing Tweet deletion rates in comparison to other regions are summarized in Table 2.

**Since 2019, Hong Kong users have a higher rate of protecting accounts, as well as deleting Tweets, than the average Twitter user.** Of a sample of all Tweets in 2019, users that are self-reported to be based in Hong Kong are 33% more likely than the average Twitter user to have protected their account. Of these users that are still public, they are over 247% more likely to have deleted Tweets than the average Twitter user. The rates of Tweet deletion and account protection since the national security law passed are significantly above average churn on the website. These trends hold in comparison to Twitter users who are from New York City and Tokyo.

We calculate statistical significance of these figures using a two-sided binomial test. This statistical test is best-effort as it presumes independence, but Tweets are certainly not independent. We calculate the proportion of deleted accounts, protected accounts, and inaccessible tweets between our Hong Kong dataset and control dataset and find that the differences in all of these rates are statistically significant under the independence assumption. We measured a p-value of 1.74e-48 or less for the difference in user account protection and Tweet deletion and inaccessibility rates.

**Compared to the other two rates, Hong Kong users are not deleting their accounts much more than the average.** This implies that relative to the average Twitter user, users are staying online, but opting to instead protect their accounts or delete individual Tweets. We measured a p-value of 0.00127 for this statistic, which is still statistically significant, but much less so than the other figures.

**More inaccessible Tweets by Hong Kong or Taipei users are politically sensitive in nature than in other populations.** As shown in Table 3, we find that of the total inaccessible Tweets, Tweets containing "sensitive" keywords are more likely to be deleted or protected by both Hong Kong and Taipei users. We do not include the control in this analysis as the total number of "sensitive" Tweets to analyze was too small (34). We also note that the political sensitivity keywords are calibrated for the Hong Kong political context; these keywords will not capture many types of political political sensitivity in other areas.

**Our Taipei user dataset shows similar user and Tweet deletion trends as Hong Kong.** Though Taipei users are more likely to protect their accounts rather than delete them, other trends related to Tweet deletion, and especially sensitive Tweet deletion, are higher than our control and other datasets. Since national security law penalties technically extend to speech outside Hong Kong and China, Taiwanese citizens who may need to travel to China may be wary of discussing "sensitive" political issues online [27]. In 2017, for instance, a Taiwanese activist transiting through Macau was detained [8].

### 4.2 RQ2: Changes in discussion trends

The primary results from our longitudinal analysis appear in Figure 2. Overall, using our sampling method of curating groups of users from historical snapshots, users from these samples generally Tweet less over time. We normalize the statistics for the rest of the keyword analyses, which are displayed relative to the number of total Tweets in the dataset per bucket of time. The histogram charts are calculated using weeks as the unit of time.

**Tweets about sensitive political topics were increasing up until the national security law passed, after which they have been in steady decline.** This trend appears in Figure 2a. The first peak of Tweets in this graph, around August 2019, correlates well with the peak of the Hong Kong anti-extradition protests. The Tweet trends also reflect the protests subsiding around the end of 2019. Tweets increase in the summer of 2020 again before very quickly dropping after July 2020. As contrast, in our control analysis, related Tweets (generally about China) peak more around the start of the coronavirus pandemic, which originated in China.

After enactment of the NSL, in Hong Kong, these trends decline steadily and never return to prior levels. Unlike the Hong Kong users dataset, in our control dataset, these trends sometimes return to previous levels. For instance, in the control dataset, relevant activity surges around August 2021. This activity resurges again in March 2022, around when Russia first invaded Ukraine. The control data is provided for reference, but we note again that Tweets that are "sensitive" in the context of the NSL comprise significantly fewer Tweets than in the Hong Kong dataset, so the control data may be noisier.

**The relative prevalence of Tweets about other non-sensitive topics is relatively steady throughout this time frame.** Figure 2b demonstrates this trend. This shows that Hong Kong users have overall not changed their Tweeting frequency when it comes to topics that are likely not covered by the national security law, such as travel, food, art, and media discussion. Our control group shows the same trend.

**After an initial peak of Tweets mentioning the Coronavirus, Tweeting trends in Hong Kong continued to decline despite local transmission levels.** This trend appears in Figure 2c. This is unexpected because Hong Kong has had several large COVID-19 waves, peaking dramatically in March 2022. In our control data, Tweeting activity around COVID-19 has generally been strongly correlated around local transmission rates after the peak of initial COVID discussion. The COVID-19 trend observed in our Hong Kong Tweet data is thus quite abnormal since Tweeting activity did not resurge around March 2022, and in fact continued to decrease.

## 5 Conclusion

In this work, we show that since enactment of the national security law, Hong Kong users demonstrate two types of online self-censorship. First, these users are more likely than other populations to remove or restrict access to their past online activity. Second, these users are posting less often about politically sensitive topics while continuing to discuss other topics online.

Taken in conjunction with related work, such as the analysis of Turkish Twitter users by Tanash et al. [23], our case study reinforces the dramatic effect that local public policy and law can have on self-censorship of online speech.

One direction for future work includes performing more specific short-form topic modeling experiments than keyword trends. This analysis might provide more nuanced understanding of individual self-censorship de-

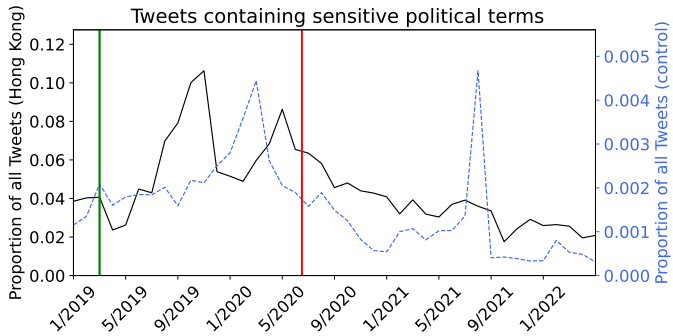

(a) Tweets containing "sensitive" keywords, compiled using keywords censored on Chinese chat apps.

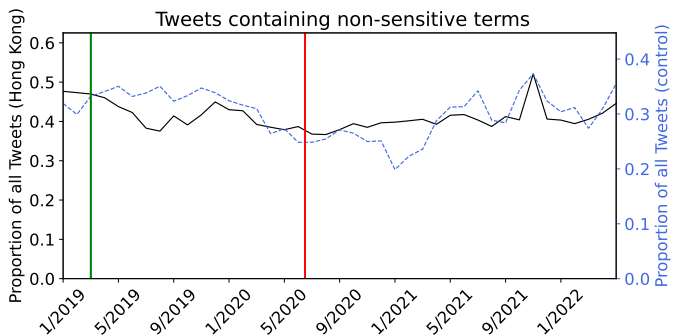

(b) Tweets containing "nonsensitive" keywords, including terms and hashtags related to media, art, music, and travel that were popular in each dataset.

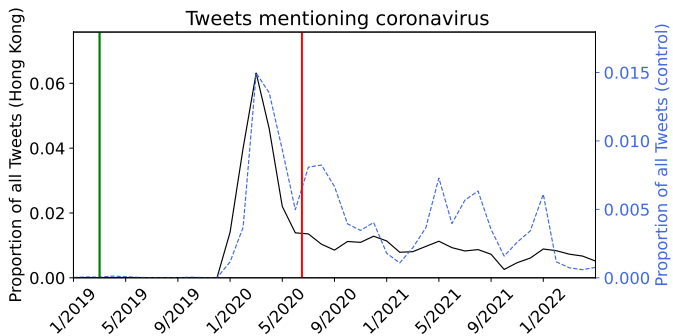

(c) The proportion of Tweets discussing COVID

Figure 2: Analysis of particular sets of keywords over time, as Tweeted by Hong Kong users and a control group. The first green line marks mid-March 2019, when the anti-extradition protests first started. The red line in the center marks enactment of the national security law.

cisions. Sentiment analysis of Tweets could also supplement this data, demonstrating how particular viewpoints on politically sensitive topics may relate to self-censorship.

Another avenue for future work could include examining different or narrower categories of online users. For instance, do journalists and academics change their behavior differently from other online users in response to laws that impose penalties for speech? Do these trends change for users who Tweet in particular languages?

In anticipation of future policy events that might influence online speech, researchers can use our tooling and methodology to study large-scale samples of Tweet streams. For instance, we might suspect similar online trends to unfold for the upcoming Turkish law that makes individuals, rather than platforms, liable for the spread of disinformation [1]. As more and more countries attempt to restrict online speech, we expect additional empirical evidence of large-scale self-censorship to emerge in response to these policies.